# A Fourier-Based Global Denoising Model for Smart Artifacts Removing of Microscopy Images

Huanhuan Zhao[1], Connor Vernachio[2,3], Laxmi Bhurtel[2,3], Wooin Yang[2], Ruben Millan-Solsona[4], Spenser R. Brown[5], Marti Checa[4], Komal Sharma Agrawal[5], Adam M. Guss[5], Liam Collins[3], Wonhee Ko[2,3*], Arpan Biswas[3,6*]

[1]Bredesen Center for Interdisciplinary Research, University of Tennessee, Knoxville, USA, 37996
[2]Department of Physics and Astronomy, University of Tennessee, Knoxville, Tennessee 37996, USA
[3]Center for Advanced Materials and Manufacturing, University of Tennessee, 2641 Osprey Vista Way, Knoxville, Tennessee 37920, USA
[4]Center for Nanophase Materials Sciences, Oak Ridge National Laboratory, Oak Ridge, USA, 37830
[5]Biosciences Division, Oak Ridge National Laboratory, Oak Ridge, USA, 37830
[6]University of Tennessee-Oak Ridge Innovation Institute, University of Tennessee, Knoxville, USA, 37996

**Email**: wko@utk.edu, abiswas5@utk.edu

*Microscopy such as Scanning Tunneling Microscopy (STM), Atomic Force Microscopy (AFM) and Scanning Electron Microscopy (SEM) are essential tools in material imaging at micro- and nanoscale resolutions to extract physical knowledge and materials structure-property relationships. However, tuning microscopy controls (e.g. scanning speed, current setpoint, tip bias etc.) to obtain a high-quality of images is a non-trivial and time-consuming effort. On the other hand, with sub-standard images, the key features are not accurately discovered due to noise and artifacts, leading to erroneous analysis. Existing denoising models mostly build on generalizing the weak signals as noises while the strong signals are enhanced as key features, which is not always the case in microscopy images, thus can completely erase a significant amount of hidden physical information. To address these limitations, we propose a global denoising model (GDM) to smartly remove artifacts of microscopy images while preserving weaker but physically important features. The proposed model is developed based on 1) first designing a two-imaging input channel of non-pair and goal specific pre-processed images with user-defined trade-off information between two channels and 2) then integrating a loss function of pixel- and fast Fourier-transformed (FFT) based on training the U-net model. We compared the proposed GDM with the non-FFT denoising model over STM-generated images of Copper (Cu) and Silicon (Si) materials, AFM-generated Pantoea sp. YR343 bio-film images and SEM-generated plastic degradation images. We showcased the tuning effect between two imaging input channels in trading-off performance between artifacts removal vs feature preservation. We believe this proposed workflow can be extended to improve other microscopy image quality and will benefit the experimentalists with the proposed design flexibility to smartly tune via domain-experts' preferences.*

High-resolution microscopy is a cornerstone of scientific discovery, providing unparalleled insight into the structure and function of different material properties and biological specimens. However, a fundamental challenge in microscopy is the complexity in acquiring high image quality. This is required for extracting fine structural details and is particularly problematic in material science fields where significant hours are spent to tune the microscopy operations for standard image quality. Despite the rapid advancement of autonomous and AI-assisted microscopy, the automation of scanning probe microscopy and electron microscopy remains constrained by persistent challenges in image fidelity and noise management. The inherent sensitivity of these techniques to environmental and instrumental perturbations introduces significant variability in data quality, which poses a bottleneck for automated image interpretation and feedback-driven control. In scanning tunneling microscopy (STM), the tunneling current is exponentially dependent on the tip–sample separation. Consequently, pronounced fluctuations in measured current can occur even with tiny vibrational and electrical noise, or environmental changes like temperature drift, leading to artifacts (e.g. scan lines, periodic height fluctuation, tip jump) and distorted surface features[1–3]. Similarly, in atomic force microscopy (AFM), cantilever deflection measurements are influenced by the mechanical response of the tip–sample interaction, as well as by acoustic or thermal noise from the environment and nonlinearities in the electronic feedback system. This dependence on system behavior and the image acquisition process not only reduces contrast and generates artifacts (e.g., line drift, tracking artifacts, scanning discontinuities), but also introduces inconsistencies across repeated scans, limiting the performance of pattern recognition and image analysis methods based on machine learning, making it necessary to account for such artifacts during the training of deep learning models[4]. Moreover, these effects degrade the performance of data-driven models and affect the accuracy of large-area image stitching[5], highlighting the importance of developing denoising approaches to improve analysis and reconstruction workflows[6]. Similarly, obtaining non-noisy images in Scanning Electron Microscopy (SEM) remains challenging due to the interplay of beam conditions, sample properties, and environmental factors. Low beam currents or short dwell times reduce signal-to-noise ratio, while high currents can cause surface charging or beam damage. Non-conductive or rough samples introduce charging artifacts, and conductive coatings used for mitigating this may obscure fine features. Additional noise arises from detector sensitivity, vibration, or vacuum instability. These issues make it difficult to achieve consistent image quality, posing limitations for automated and machine-learning-based SEM analysis and feature extraction[7,8]. Thus, automated microscopy systems rely heavily on consistent, high-quality image datasets for feature detection, phase identification, and learning material structure-property relationship. The presence of such stochastic noise and non-physical patterns—originating from tip wear, surface contamination, or feedback instability—can lead to misclassification or convergence failures during autonomous exploration.

Several researches have been conducted in an attempt to improve the quality of microscopy images through different denoising models such as from supervised to unsupervised and from data-driven to physics-driven[9,10]. For example, Joucken et. al.[2] has developed a supervised CNN-based denoising model, trained on a set of STM simulated images based on a tight-binding electronic structure model. Sadri et. al.[11] demonstrates an unsupervised deep learning model to denoise 4D STEM data. Xie et. al.[3] developed an unsupervised physics-augmented adversarial domain adaptive model (PDA-Net) to denoise STM images, trained over simulated images. Chang et. al.[12] proposed a two-stage denoising method for SEM images where stage 1 employs the variance stabilization to separate the signal-dependent noise and stage 2 employs a U-net based noise removal mechanism. Tian et. al.[13] proposed a zero-shot self-supervised learning denoising framework for High-Resolution Electron Microscopy (HREM) using a single noisy HREM image, which helps to generate approximate infinite noisy pairs for training. Xio et. al. developed a SEM image denoising model based on the denoising diffusion probabilistic models (DDPM). Li et. al.[14] compared 16 different denoising models to remove stripe artifacts in Conductive atomic force microscopy (c-AFM) images, and concluded Low-Rank Recovery method has the best performance. Kocur et. al.[15] design a ResU-Net architecture to denoise AFM image artifacts, trained over synthetically generated data. Zhang et. al.[16] developed a method based on Bayesian compressed sensing to remove impulse noise from atomic force microscopy (AFM) images. However, these methods are either supervised learning, or need large set of simulated data, or focused on removing artifacts, or focused on particular microscopy operations. Furthermore, the complexity of noise characteristics in different microscopies makes it difficult to design universally applicable denoising or image reconstruction algorithms without compromising real structural information[3]. In materials science for property discovery (e.g. quantum phenomenon), it is not always known which signals are noise vs desired properties, which restricts any supervised learning. For example, interference of scattered electrons and tip-height fluctuation both generate periodic wavy pattern in STM images, where the former is a desired feature, but the latter is an artifact[3,17]. In other cases, such as in medical diagnostic, studying bacterial biofilm formation[5] and material degradation, simulating images is not an option and need to rely only on computationally costly and noisy experimental training images. On the other hand, the definition of artifacts and features vary on different materials and different microscopy, which limits the integration of a global physical knowledge into the denoiser model.

To summarize, in this age of automated microscopy, a low-cost ML-driven denoised high quality data acquisition is required in order to accurately post-process the images for meaningful physical discovery and can be applicable across various sources of microscopy. One way to support such low-cost training is the use of minimal training images and cheaply acquired noisy images. Such denoised algorithm, named as Noise2Noise[18], has been developed where multiple noisy image pairs are used for training. However, the acquisition of multiple noisy image pairs is not often feasible and therefore has been further developed to Noise2Void (N2V)[19] which requires only a single noisy image. N2V has been further upgraded to N2V2[20] and Noise2Fast (N2F)[21] models in order to improve performance and computational efficiency. Inspired from the N2V, we present a global denoising model (GDM) to smartly improve quality of microscopy images which aim to 1) consider only two non-pair simulated or experimental training images as available (low data without dependence on clean images), 2) learning microscopy and material system specific artifacts during training to preserve other strong- and weak- signaled features (goal-specific), 3) adding flexibility to tune the training trade-offs between artifact removal and feature preservations (domain expert preferences). Once we train the proposed goal-specific denoise model with domain-specific tuning, the GDM can be utilized to either offline denoise pre-acquired experimental images or online denoise the experimental images in real-time microscopy data acquisition. To our knowledge, only Corrias et. al.[22] recently attempted to develop a total variation-based image decomposition and denoising model and applied for a large variety of microscopy such as AFM, STM and SEM. In this paper, we designed the denoised model based on the deep learning U-net model with developing an integrated pixel- and fast Fourier-transformed (FFT) loss function. U-net architecture has been adopted significantly in recent years in image denoising[23–26] and has been extended to denoise microscopy images[27]. **Figure 1.** shows the proposed architecture of the GDM.

Here, we first choose two non-pair images, which can be either simulated images or experimental images or both as per availability for the denoising of other experimental images from similar systems. In Fig. 1, we provided the example of Cu(111) images where input channel 1 is the experimental image and input channel 2 is the simulated image. Details of the generation of these images are provided in the result section. Then based on the domain-specific knowledge, we set the goal to pre-process the data to enhance the known features. In this example, the black dots and the surrounding concurrent circles on the black dots are the features which explain the defects (sulfur or oxygen atoms) and quantum interference patterns of the surface-state electrons. On the other hand, the continuous zig-zag lines throughout the experimental images are the artifacts of STM scan lines. Then these two images are converted into set of patches, with user specified patch size, to increase the volume of training dataset. Next, we create random masking for each patch of each of the training images, with user specified fraction of the masking. Then, we fit a deep learning compact U-Net–based convolutional neural network following with an encoder–decoder structure with symmetric skip connections that facilitate the recovery of fine-grained spatial details lost during down-sampling. Each convolutional block in the network consists of two consecutive 3×3 convolutional layers (padding = 1), each followed by a rectified linear unit (ReLU) activation. The encoder path comprises three convolutional blocks with progressively increasing channel depths of 32, 64, and 128 respectively and using 2×2 max pooling between successive encoder stages. The decoder path employs bilinear up-sampling (scale factor = 2) to progressively restore spatial resolution. At each decoding stage, the upsampled feature maps are concatenated with the corresponding encoder outputs via skip connections, ensuring the retention of high-frequency spatial information. The decoder includes two convolutional blocks, reducing the feature dimensions from 128 to 64 and finally to 32 channels. A final 1×1 convolutional layer maps the output of the last decoder block to a single-channel image, matching the dimensionality of the input. We train this U-Net architecture with the generated masked images patches from two input channels, in batches with user specified batch size, and predict the output of the masked region. Then we define our proposed loss function which consists of 1) mean square error (MSE) between the ground truth, $ip_{,1}$, and the predicted output, $\overline{ip_{,1}}$, in the masked region and 2) cosine similarity between the FFT of the ground truth, $ip_{,1}$, and the predicted output, $\overline{ip_{,1}}$. Then, each of the patches within each batch of the two input channels, $c_1$ and $c_2$ are evaluated as mentioned. Then, we computed the weighted average pixel loss, $l_{px}$, and FFT loss, $l_{FFT}$ for all the patches per batch, $B$, in both input channels with the user-defined weighting factor of $w_1$ and $w_2$ respectively, as provided as per **Eq. 1-2**. Finally, we calculate the weighted total loss, $L$, as per **Eq. 3**. We train the U-net model batchwise (to reduce computational load) by minimizing the weighted total loss via Adam optimizer with user-specified learning rate and decay by a factor of 0.5 after every 10 epochs.

$$l_{px} = w_1 * \sum_{b=1}^{B} MSE(ip_{c1,1}, \overline{ip_{c1,1}}) + w_2 * \sum_{b=1}^{B} MSE(ip_{c2,1}, \overline{ip_{c2,1}}) \tag{1}$$

$$l_{FFT} = w_1 * \sum_{b=1}^{B} \Psi_{FFT}(ip_{c1,1}, \overline{ip_{c1,1}}) + w_2 * \sum_{b=1}^{B} \Psi_{FFT}(ip_{c2,1}, \overline{ip_{c2,1}}) \tag{2}$$

$$L = \frac{l_{px}}{1+l_{FFT}} \tag{3}$$

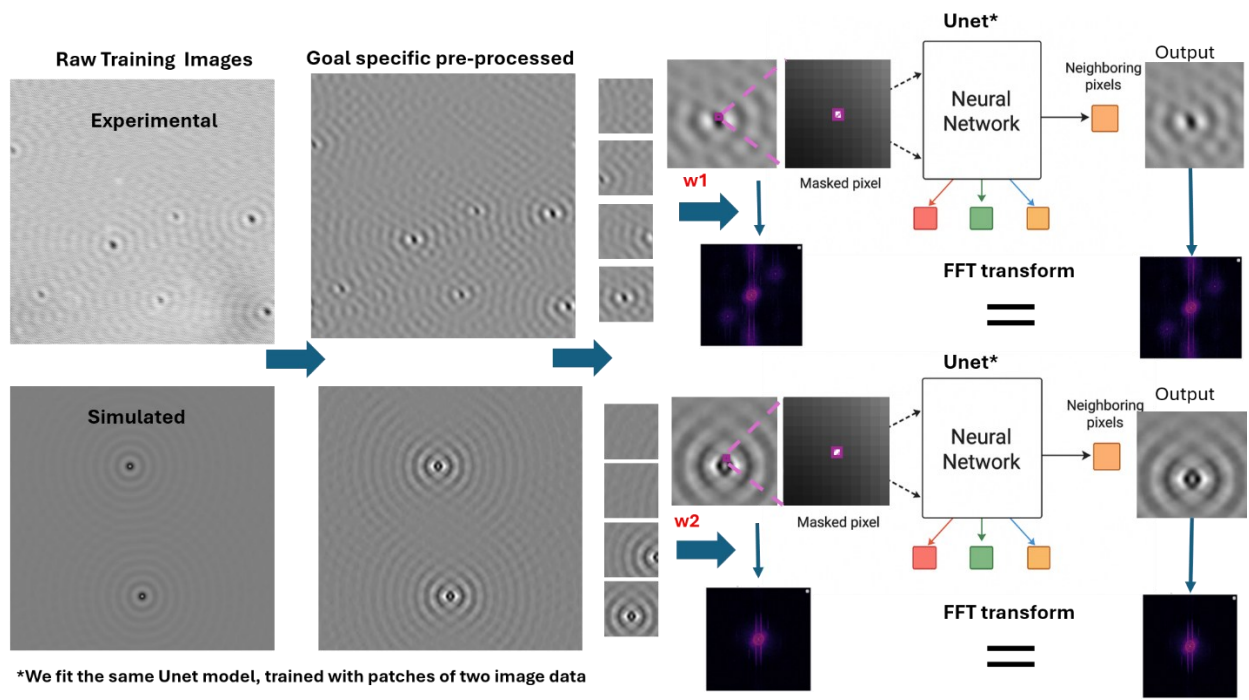

***Figure1:*** *Architecture of the proposed Fourier-Based Global Denoising Model (GDM).*

## Result and Discussion

In this section, we demonstrated the performance of our proposed GDM over various sets of pre-acquired microscopy data, focusing on different patterns of features, noises and artifacts in the images.

**Case Study 1: Denoising STM images**

Here we have presented the denoising of STM images of two material systems- 1) Cu(111) single crystal and 2) Si(111) surface with 7×7 reconstruction. Both Cu(111) and Si(111) are prototypical materials that have been extensively studied by STM. A single crystal Cu (111) was prepared in an ultra-high vacuum (< $5×10^{-10}$ torr) chamber by several cycles of annealing and sputtering to get atomically flat and clean surfaces. Each cycle consists of $Ar^+$ ion sputtering at 800 V for 10 minutes followed by annealing at 500°C for 30 minutes. Then, Cu(111) was transferred to the low-temperature STM head at 4 K for scanning, where topographic images were acquired with bias voltage of 10 mV, setpoint current of 1 nA, scan size of 30×30 $nm^2$, and pixel number of 512×512 [Fig. 1, Fig. 2(a)]. The surface exhibits standing waves patterns originated from the interference pattern of electrons scattered by atomic defects, also known as quasi particle inference (QPI) pattern[28]. Simulated images of Cu(111) with QPI pattern were generated by Green's function method following Fiete *et al* [29]. In Fig. 1, the simulated image (scan size of 35×35 $nm^2$, pixel number of 512×512) displays the calculated local electron density of states (LDOS) at the Fermi level of Cu(111) surface states, which were modeled as two-dimensional free electron gas with effective mass of $0.38m_e$ and chemical potential of 0.45 eV, where $m_e$ is the bare electron mass. A clean Si(111) surface with 7×7 reconstruction was prepared by repeated flashing in a UHV chamber. A Si(111) wafer (p-type, 0.001 Ω·cm) was outgassed at 500°C for several hours. Then, the sample was flashed to 1150°C for 5 s to remove surface oxides, and subsequently to 950 °C for 5 s to form 7×7 reconstruction. After flashing, the sample
6

gradually cooled down to room temperature. The prepared sample was transferred to the STM head at 4 K to acquire topographic images with bias voltage of 2 V and setpoint current of 100 pA (scan size of 10×10 nm$^2$, pixel number of 128×128 in Fig. 5). Simulated images of Si(111) surface with 7×7 reconstruction was generated by Nanonis Mimea STM simulator, SPECS Zurich GmbH, (scan size of 20×20 nm$^2$, pixel number of 256×256 in Fig. 5).

Here, during the model training, wet set the model parameters as *patch_size* = 128, *batch_size* = 8, *learning rate* = 1e$^{-4}$, *mask_fraction* = 0.1, *epoch* = 50. **Figures 2 and 3** shows the performance of the GDM, at different weighting factor of $w_1$ and $w_2$, without and with using the goal-specific pre-processed training images respectively. From figs. 2b, we can see the FFT peak intensities away from the center decreases which signifies the noise reduction. Here in the PNR calculation, we assume the noise intensities lies outside 50% of the maximum radius from the center point, in order to avoid getting higher scores for removing weaker but physically-important features. We can see the denoised images with higher weightage on simulated images as $w_2 = 0.99$ and $w_2 = 1$ provides siginificantly higher scores. This is likely as FFT loss can be better computed from from cleaner simulated images, thus minimizing the surrounding peak intensities better. Figs. (c) shows another way to compare the performance of the denoised images where we draw the detected lines over the original noisy and the subsequent denoised images. More the concentration of the thin lines represent likely to have higher artifacts (i.e. scan lines) to these images, which we aim to reduce. Also, it is to be noted the lines can also represent the feastures, and therefore just minimizing the total length of the lines might not be a feasible solution. Therfore, based on the nature of angle of STM scan lines in the range of (−30, 30), we draw the overlapped detected red lines and calculated the total length of the scan lines in that range. We can see that while the denoised image at $w_2 = 0.99$ shows the highest PNR score in terms of keep the weaker but physically important feastures (such as concurrent circles surrounding the black dots), it did not remove the noise as efficiently as we can see for the denoised image at $w_2 = 1$, as we compare the total line lengths. Visually in fig. 2c, we can see for $w_2 = 1$, black holes (representing the dots in the experimental images in figs. 2(a)) can be clearly seen in the line plots, which also suggests better noise removal but likely at the cost of blurring some weak physical features. Comparing the results from fig. 3(b), we can clearly see all PNR scores of the denoised images increased from the respective denoised images in fig. 2(b). We can see a significant reduction in the intensities of the vertical lines at the edges of the FFT images, which suggest the denoising of the artifacts. At the same time, looking into figs. 3(c), the quality of all the denoised images further improved into detecing those black holes. We can see the denoised image at $w_2 = 0.99$ shows the optimum metric in highest PNR score of 14.88 and significantly lower total line length of 6791.84. This shows with domain knowledge driven goal-specific pre-processing, we can further reduce the noisy artifacts while maximize the preservation of the feature intensities. To summarize, all the denoised images in Figs 2 and 3 show better quality in terms of visual inspection, PNR scores and estimated artifact-based total length scores than the orginal experimental images.

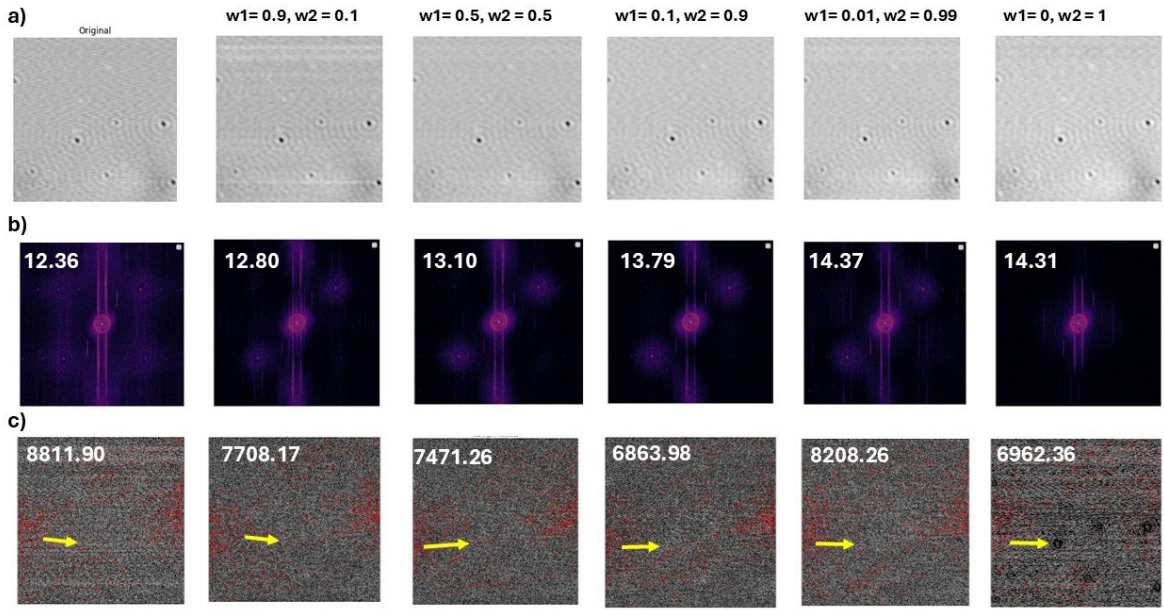

*Figure 2:* Results of the proposed Fourier-Based Global Denoising Model (GDM) over STM generated Copper Cu(111) experimental image. Here, the **raw simulated and experimental data** were directly inputted in training. The figures in (a) are the original experimental image, followed by the denoised images at different trade-offs ($w_1, w_2$) between training image 1 (experimental image in Fig. 1) and training image 2 (simulated image in Fig. 1). The figures in (b) are the respective FFT peak intensity plots with the calculated Peak-to-Noise (PNR) scores (higher the better). The figures in (c) are the detected lines plots over the respective images in (a) with the calculated total line length within the range of angle $= (-30, 30)$ as highlighted in red (lower the better). This angle range is considered as per the domain expert knowledge of potential scan lines (artifacts).

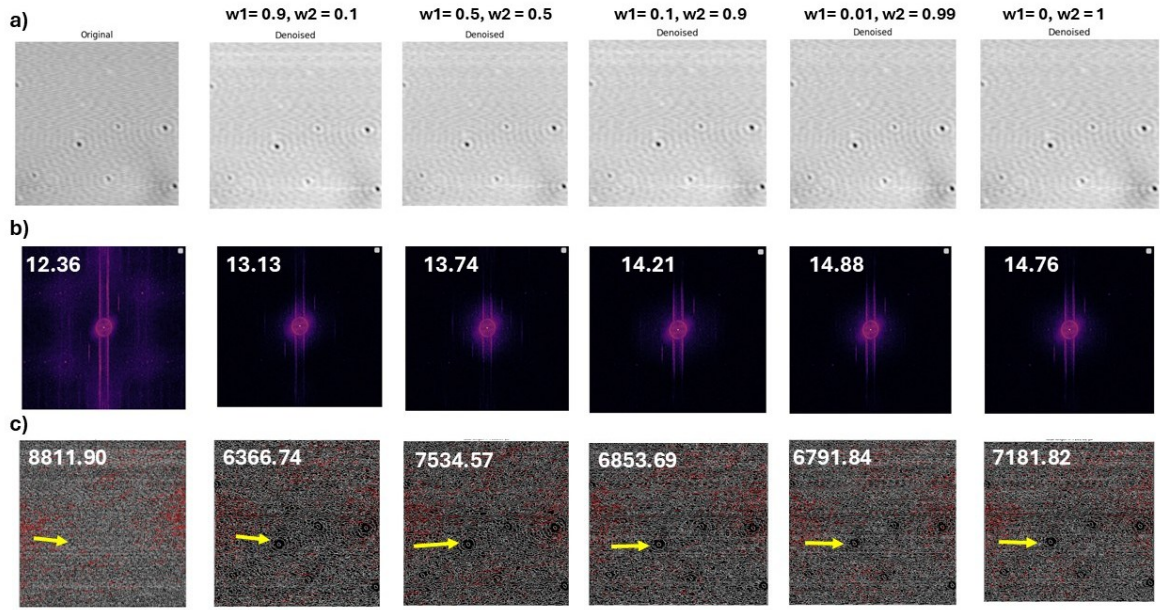

*Figure 3:* Results of the proposed Fourier-Based Global Denoising Model (GDM) over STM generated Copper Cu(111) experimental image. Here, the **pre-processed simulated and experimental data** were *directly inputted in training. The figures in (a) are the original experimental image, followed by the denoised images at different trade-offs ($w_1, w_2$) between training image 1 (experimental image in Fig. 1) and training image 2 (simulated image in Fig. 1). The figures in (b) are the respective FFT peak intensity plots with the calculated Peak-to-Noise (PNR) scores (higher the better). The figures in (c) are the detected lines plots over the respective images in (a) with the calculated total line length within the range of angle $= (-30, 30)$ as highlighted in red (lower the better). This angle range is considered as per the domain expert knowledge of potential scan lines (artifacts).*

**Figure 4** compared the earlier reported denoised image (fig. 4b), providing optimum performance metric, with other methods (figs 4 c-f) of microcopy image denoising. Compared with fig. 4c where FFT loss has not been integrated with the MSE of pixel loss, we can see new artifacts of white horizontal lines (highlighted by yellow arrow) which are clearly not present in the training simulated image or tested experimental image (fig. 4a). Compared with the denoised image from PDA-Net model[3] as in fig. 4d, we can see clearly the weaker physically relevant signals of concurrent circles are completely blurred along with the noisy scan lines and all the black dots are populated with white circled artifacts which are not present in the experimental images. Additionally, we can see in some areas (highlighted by green square region) the weaker concurrent circles are replaced by such white circled artifacts. Moreover, unlike our model, the PDA-Net model needs multiple clean images to train, which shows the advantage of our method. Compared with the denoised image from N2V model as in fig. 4e, the performance gives the closest resemblance to our model. However, looking closely at some of the regions as highlighted by the red and green squares, we can see our model has provided slightly enhanced weaker signals (weaker black dots and concurrent circles within red and green squares respectively). Thus, our model provides better trade-offs in preserving the weaker features. Compared with the denoised image from N2V2 model as in fig. 4f, we can see similar new artifacts of white horizontal lines (highlighted by yellow arrow).

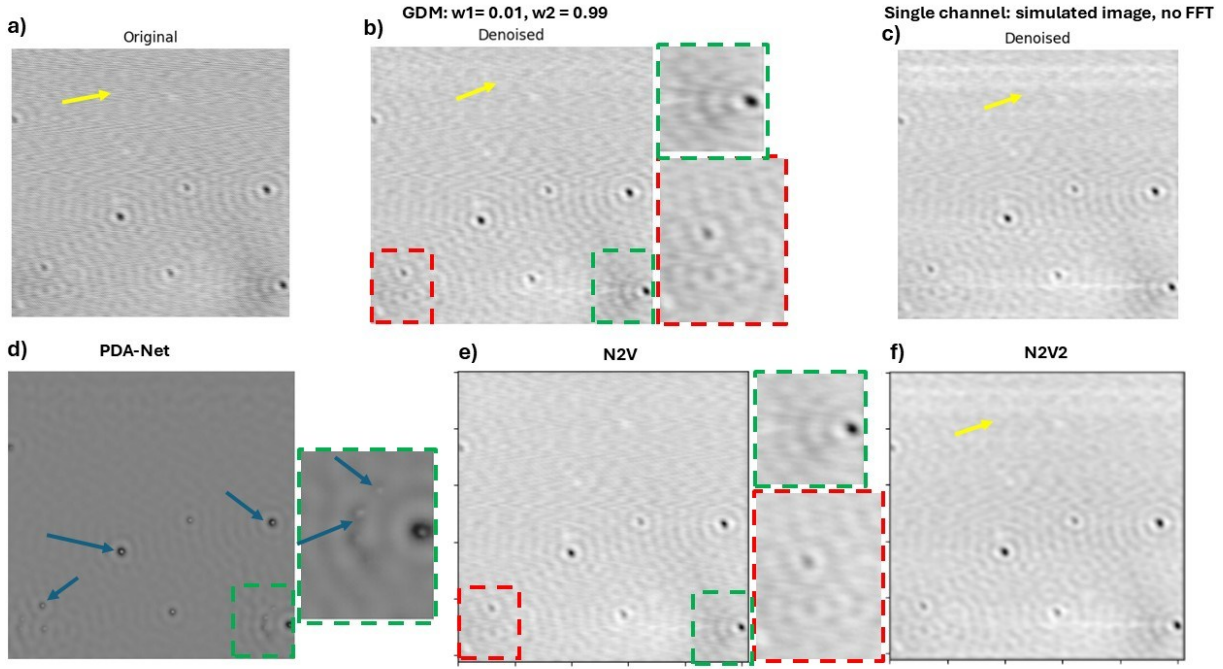

*Figure 4: Comparison of the proposed Fourier-Based Global Denoising Model (GDM) with other existing methods over STM generated Copper Cu(111) experimental image. Here, (a) is the noisy experimental image, (b) is the denoised image from the proposed GDM with the best PNR score reported in Fig. 2, (c) is the denoised image from the model without the FFT-based loss (considering only pixel based MSE loss), (d) is the denoised image from the PDA-Net model[3], (e) is the denoised image from the N2V model[19] and (f) is the denoised image from the N2V2 model[20]. For (c), (e) and (f), the preprocessed simulated image is used for training. For (d), approximately 100 simulated images of Cu(111) are used for training.*

**Figure 5** shows the performance of GDM over Silicon Si(111) image, where a single simulated image (fig. 5a) is used for training. We can clearly see the scan lines artifacts in the original noisy top and middle images of fig. 5b, which we aim to remove. On the other hand, the white particles and the hexagonal atom structure are the physically relevant features, which we aim to enhance. From the respective FFT peak intensity bottom figure of fig. 5b, we can see such scan line artifacts are populated as high intensity peaks near the edges. From fig. 5c, we can see our model significantly removed those scan line artifacts and has preserved the stated physically relevant features. From the respective FFT peak intensity bottom figure of fig. 5c, we can see those high intensity peaks near the edges have been minimized while preserving the high intensity peaks near the center, thereby increasing the PNR score from 17.95 to 21.65. Compared with fig. 5d where FFT loss has not been integrated with the MSE of pixel loss, while we can see the scan lines artifacts have been significantly removed, there are some regions (highlighted in red square) where fake defects have been introduced into the atom hexagonal structure (the enhanced dark region highlighted by yellow arrow). This addition of new artifacts can be seen in the respective FFT peak intensity figure as we can see while the high intensity peaks near those top and bottom edges have been minimized, new high intensity peaks are populated near the left and right edges. Therefore, we see a lower PNR score of 18.88 compared to the PNR score of 21.65 from our method.

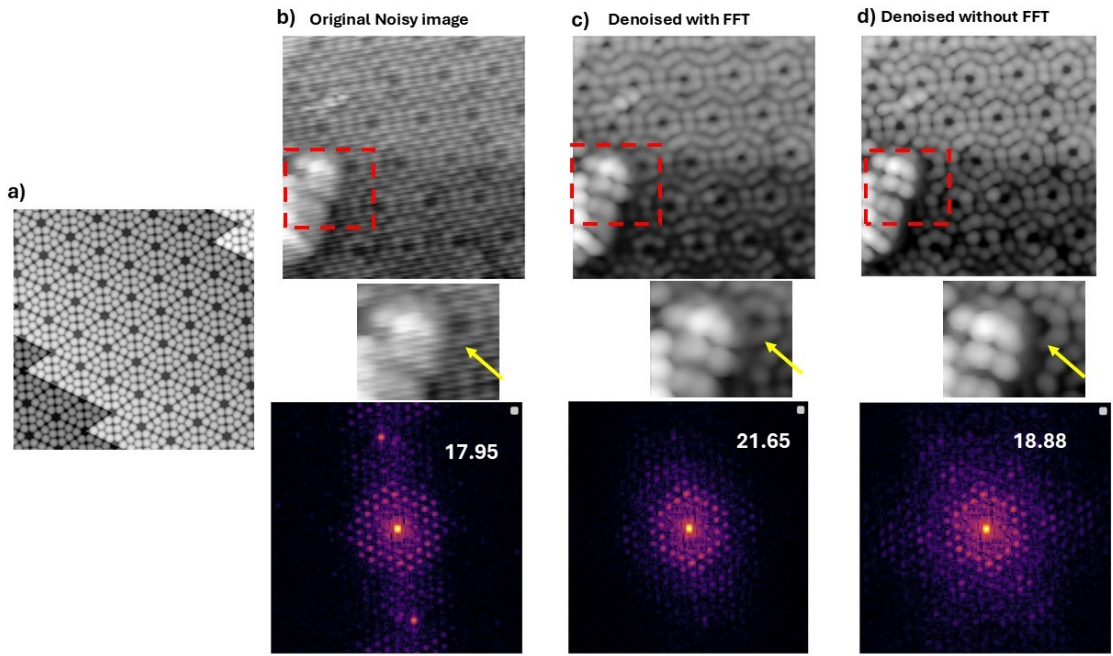

*Figure 5: Results of the proposed Fourier-Based Global Denoising Model (GDM) over STM generated Silicon Si(111) experimental image. Here, **only the raw simulated data**, as shown in figure (a) were directly inputted in training. The top and bottom figures in (b) are the original experimental image and the*

*respective FFT peak intensity plots with the calculated Peak-to-Noise (PNR) scores (higher the better). The top and bottom figures in (c) and (d) are the denoised experimental image and the respective FFT peak intensity plots from the proposed GDM and the model without the FFT-based loss (considering only pixel based MSE loss). The red dotted square in the top figures of (a)-(c) are the provided zoomed middle figures.*

**Case Study 2: Denoising AFM images**

Here, we implemented our proposed GDM to AFM generated topographical images, dimension of 15 x 15 µm$^2$, of early biofilm assembly by ***Pantoea* sp. YR343**, where the strain was grown on a silicon oxide substrate between 4 and 6 hours. AFM images were acquired using the *DriveAFM* system from *Nanosurf*, which features a motorized platform capable of scanning large areas. The datasets were collected using the methodology described in[6] through a custom Python graphical interface designed to capture image arrays. More details on the sample preparation and dataset generation can be found in [5,30]. AFM topographic images often contain noise, such as artificial horizontal lines caused by the feedback system, adhesion artifacts, or changes in resonance frequency, which result in abrupt variations between consecutive scan lines. Here, we aim to remove these artificial scan lines via GDM, using a minimal number of experimental images and without the need for simulated data. To achieve this, we selected a single representative experimental image and applied a preprocessing pipeline that generates multiple partial and complementary versions of the same image, which were then used as inputs for both channels of the model. This approach enables a self-supervised training strategy that relies solely on available experimental data, eliminating the need for synthetic images or large datasets. The main advantage of this procedure is that it allows the model to be trained effectively using only a very small number of high-resolution experimental images, significantly reducing data acquisition time, annotation effort, and computational cost while preserving the high fidelity of the reconstructed topography**.** During the model training, wet set the model parameters as *patch_size* = 128, *batch_size* = 8, *learning rate* = 1e$^{-3}$, *mask_fraction* = 0.1, *epoch* = 50.

**Figure 6** shows the pre-processing of the experimental training image and the performance of the GDM over removing the artificial scan lines from two test images. Fig. 6a shows the original experimental image with artifact scan lines reducing contrast. We first preprocessed this image into a "dark-masked image" by first using a FFT-based filtering algorithm to identify and suppress the horizontal scan lines and then uses an intensity percentile threshold based masking to separate the substrate regions with low information content (dark areas) from the regions containing features (bacteria).To keep the features of the original image in the non-dark region, we overlap the non-masked region (black region) of the "dark-masked image" with the featured grayscales of the original experimental image, to maintain the structural fidelity of the unmasked areas. We call this image as the "merged image". Details of the preprocessing of the images is provided in the Method Section. We used the merged and the dark-masked images as the input channel 1 and 2 respectively, and trained the model with tuning the trade-off parameter between the channels. During the training process, we find the best tuning parameter is set as $w_1 = 0.5$, $w_2 = 0.5$. Comparing Fig. 6b and 6c, we can clearly see the artificial scan lines over the dark regions are completely removed, while keeping the integral features. Thus, we see an improved image quality by increasing the PNR score from 9.11 to 12.65. Similarly observation can be found as we check the line plots in Fig. 6b, where those dark regions are expectedly populated with lines (representing the artifacts scan lines). Comparing with Fig. 6c, we can see those lines are not present in the respective dark regions which suggests the actual removal of the artifacts. Thus, we see a decrease in the estimated total line length within the range of angle $= (-10,10)$ from 3718.81 to 2274.52. Comparing Fig. 6d and 6e with another test image, we see similar observation which signifies the performance of the GDM.

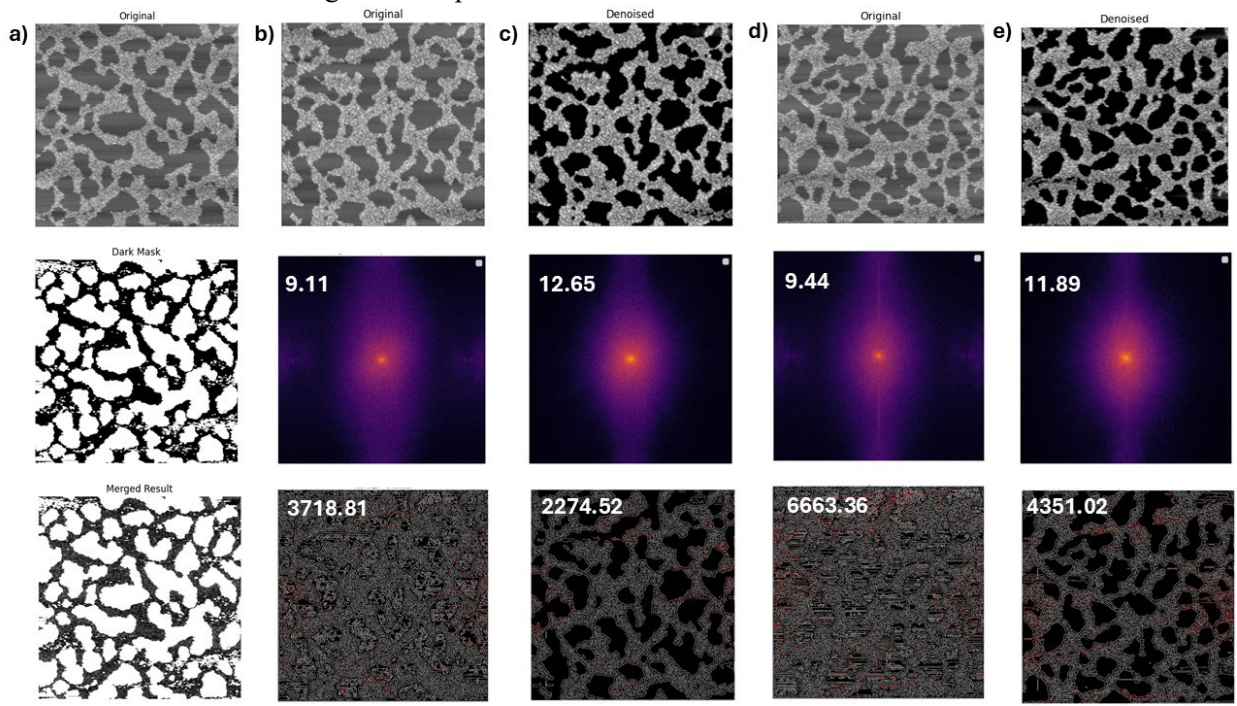

*Figure 6: Results of the proposed Fourier-Based Global Denoising Model (GDM) over AFM generated **Pantoea* **sp. YR343** *biofilm experimental images. Here, the **pre-processed experimental data** were directly inputted in training. The top, middle and bottom figures in (a) are the original, masked and merged training experimental images. The merged and the masked images are inputted into the GDM as image channel 1 and 2 respectively. The top, middle and bottom figures in (b) and (c) are the respective original and denoised test image 1, followed by FFT peak intensity plots and detected line plots with the calculated total line length within the range of angle $= (-10,10)$ as highlighted in red. The top, middle and bottom figures in (d) and (e) are the respective original and denoised test image 2, followed by FFT peak intensity plots and detected line plots. FFT peak intensity plots with the calculated Peak-to-Noise (PNR) scores.*

To understand the effect of tuning $w_1$ and $w_2$ and the integration of FFT loss, we compare the GDM with different weights of the input imaging channel and without the FFT loss integration as per **figure 7.** Compared with fig. 7b without the FFT loss integration, under the optimal weighting trade-off $w_1 = 0.5$, $w_2 = 0.5$, we can clearly see the significant presence of those artifact horizontal scan lines over the dark regions. We see the same observation as we check the GDM model training with higher weightage of information from the training merged image (figs. 7c-d). This could be due to the fact while the merged image is still masked (no artifacts) over the dark regions, it still contains the artifacts hidden in the non-dark region with the features. Thus, with more weightage of the merged image, the model starts to learn about the artifacts (in non-dark regions) as features and therfore does not remove the same efficiently from the dark-region of the test image. It is to be noted, the preprocessing of removing artifacts in the non-dark region is extremely complex as it can also remove the features as can be seen from the dark-masked training image. As we compare with giving higher weightage of information on the dark-masked image (figs. 7e-f), we see the efficient removal of the artifact scan lines. However, it compromises the preservation of the feastures. From fig. (g), as we compared with the zoomed in section of the denoised images from equal trade-off and higher weightage of dark-masked image (highlighted in red, yellow and orange borders respectively) with the orginal image (no border), we see the best matching intensity of the grayscale features with the equal weightage. Interestlingly, as we train the model with full weigtage of the dark-masked image, the denoised image started to provide several small fake dark regions (artifacts), thereby losing completely the structural integrity of the non-dark region of the noisy experimenal image. Thus, we can see with balanced source of information, the GDM model is learning optimally and therfore removed the artifacts while maximizing the feature preservation.

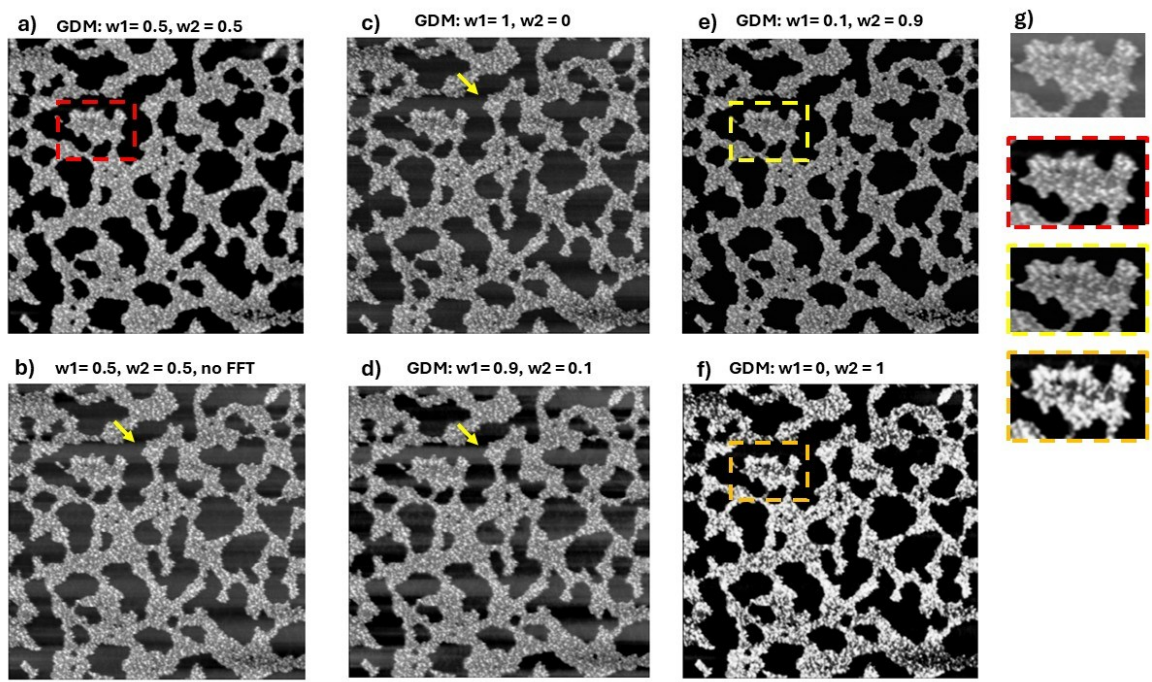

*Figure 7:* Comparison of the proposed Fourier-Based Global Denoising Model (GDM) over AFM generated **Pantoea sp. YR343** biofilm experimental image. Here, (a) is the denoised image as provided in Fig.6c with the trade-offs ($w_1 = 0.5, w_2 = 0.5$) between training image 1 (experimental merged image) and training image 2 experimental masked image), (b) is the denoised image from the model without the FFT-based loss (considering only pixel based MSE loss), (c)-(f) are the denoised image from the GDM model at different trade-offs of ($w_1, w_2$), (g) is the zoomed in comparative view of the section of the denoised images as per the highlights with the original test image (no highlight).

**Case Study 3: Denoising SEM images**
In this case study, we implemented the GDM to SEM images of plastic material degradation mediated by bacterial enzymatic activity. Polyethylene terephthalate (PET) powder was incubated with engineered microbial strains secreting PET-degrading enzymes. Following 96 hours of cultivation at 55°C, the cultures were centrifuged, and the treated PET powder was collected. A small portion of the recovered material was imaged using a Zeiss Merlin field emission scanning electron microscope (FE-SEM) operated in SE2 mode, with an accelerating voltage of 1 kV and a beam current of 100 pA. In the SEM images, one such artifact is the artificial white/bright spot due to distortion. It is to be noted, as we mentioned earlier in Fig.5, similar white spots in the STM generated Si images were physical features. As it is difficult to purely generalize a denoising model in categorizing such features vs noise over different material systems and microcopy, thus our approach of goal-specific pre-processing of training images is beneficial. Also, like case study 2, it is not feasible to generate simulated images and therefore, we can only rely on noisy experimental training images. The goal for denoising and removing the artifacts from these images is to improve the accuracy of the calculation of physical properties (such as pit locations, shape, size, area etc.) to understand the rate of material degradation, an essential step for automated microscopy analysis and rapid physical discovery. During the model training, wet set the model parameters as *patch_size* = 128, *batch_size* = 8, *learning rate* = 1e$^{-3}$, *mask_fraction* = 0.1, *epoch* = 50.

**Figure 8** shows the pre-processing of the experimental training image and the performance of the GDM over denoising two test images. Fig. 8a shows the original experimental image with artifact white spots. We first preprocessed this image to identify bright spots using threshold method (refer to the white region in middle fig. 8a) and then using an inpainting algorithm to reconstruct the captured bright spots with the intensity of neighbouring pixels (refer to bottom fig. 8a). We call this image as the "cleaned image". Details of the preprocessing of the images is provided in the Method Section. We used the single cleaned image as the input channel to train the model. Also, this training image was captured at the magnification of 10.39 KX. The first test image, as in fig. 8b, was captured at similar maginifcation parmaeter of 10.50 KX. Comparing Fig. 8b and with the denoised image in fig. 8c, we can clearly see the FFT peak instensities at the edges of the denoisied image are completely supressed. Thus, we see an improved image quality by increasing the PNR score from 6.21 to 9.56. Similarly observation can be found as we check the line plots in Fig. 8b, where those dark pits regions are populated with several noisy lines. Comparing with Fig. 8c, we can see those lines are not present in the respective dark regions (highlighted by yellow and red arrows) with more accurate structural representation of the pits. The second test image, as in fig. 8d, was captured at a different maginifcation parmaeter of 1.0 KX. Also, this test image is distorted with white spot artifacts is large area. Comparing fig. 8d and with the denoised image in fig. 8e, we see the denoised image remove the artifacts along with the random noise. Thus, we see a higher increment in PNR score from 7.31 to 12.64. Here also, we see a better structural representation of the pits (highlighted by yellow and red arrows) from the line plots of the denoised image.

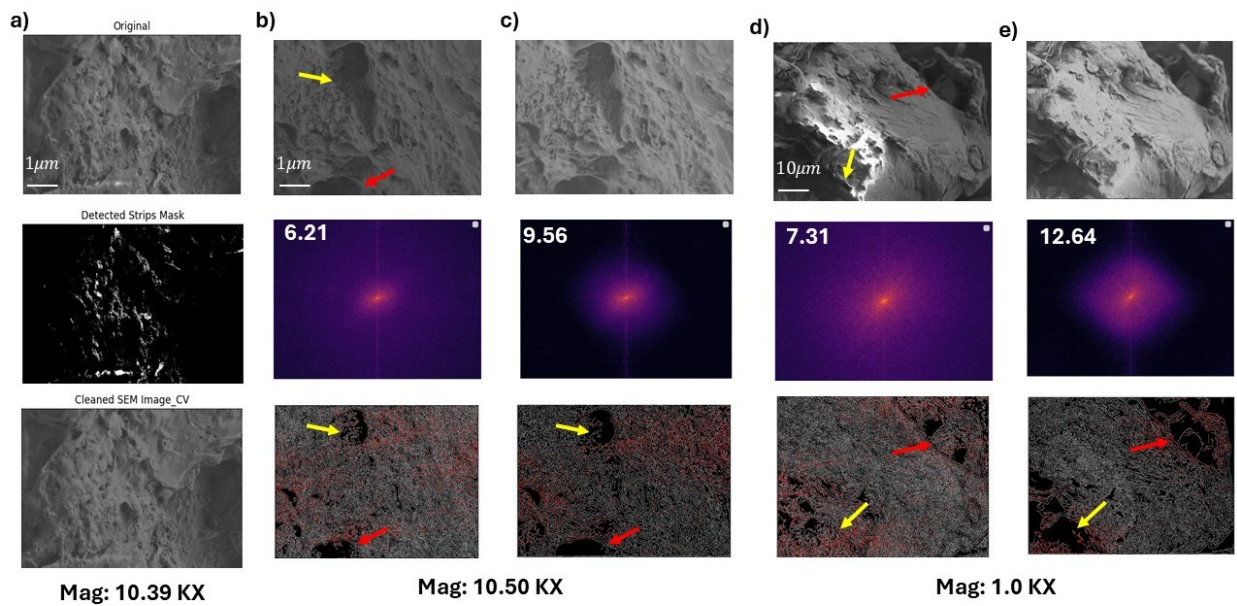

*Figure 8: Results of the proposed Fourier-Based Global Denoising Model (GDM) over SEM generated bacterial plastic consumption experimental images. Here, the **pre-processed experimental data** were directly inputted in training. The top, middle and bottom figures in (a) are the original, detected strip-masked and cleaned training experimental images. The cleaned image is only inputted into the GDM. The top, middle and bottom figures in (b) and (c) are the respective original and denoised test image 1, followed by FFT peak intensity plots and detected line plots. The top, middle and bottom figures in (d) and (e) are the respective original and denoised test image 2, followed by FFT peak intensity plots and detected line plots. FFT peak intensity plots with the calculated Peak-to-Noise (PNR) scores. It is to be noted, the training image was taken at the magnification of 10.39 KX, while the first and second test images in (b) and (d) were taken at the magnification of 10.50 KX and 1.0 KX respectively.*

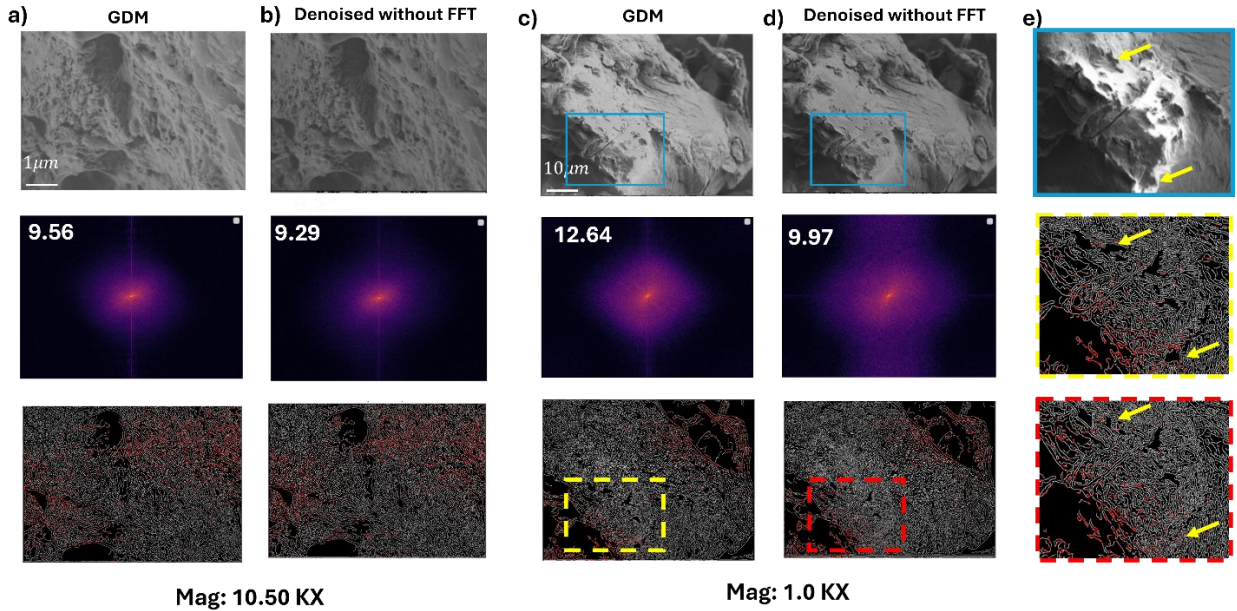

*Figure 9: Comparison of the proposed Fourier-Based Global Denoising Model (GDM) over SEM generated bacterial plastic consumption experimental images. Here, (a) is the denoised image as provided in Fig.8c, (b) is the respective denoised image from the model without the FFT-based loss (considering only pixel based MSE loss), (c) is the denoised image as provided in Fig.8e, (d) is the respective denoised image from the model without the FFT-based loss, (e) is the zoomed in comparative view of the section of the line plots of the denoised images in (c) and (d) as per the highlights with the same region of the original test image (no highlight).*

To understand the effect of the integration of FFT loss, we compare the GDM without the FFT loss integration as per **figure 9.** Compared with fig. 9a and 9b, with and without the FFT loss integration, we can clearly see that our proposed approach provided slightly better denoised image quality with PNR score of 9.56 than the non-FFT training with PNR score of 9.29. However, compared with fig. 9c and 9d, we can see a significant improvement in the performance of GDM to denoise the image (PNR score of 12.64 vs 9.97), taken at different magnification parameter. Though the non-FFT training was able to remove the white spot artifacts, it likely added some random artifacts during reconstruction. As we observe fig. 9e over the zoomed in view of the artifact's region (highlighted yellow arrows in top image of fig. 9e), we can see during reconstruction the GDM model is able to better preserve the structural integrity of the small pits (highlighted yellow arrows in middle image of fig. 9e) rather than the non-FFT loss model (highlighted yellow arrows in bottom image of fig. 9e).

## Summary

In summary, we proposed a Global Denoising Model (GDM) to improve the image quality of different microscopy images via domain-knowledge driven goal-specific pre-processing of the training images, integrating an FFT-based loss function and/or tuning weights of information from two imaging channels during training. The structural framework of the GDM is designed on a lightweight U-Net model with only 1-2 training images and no dependencies of simulated or pair images. The goal-specific pre-processing of the training images provides better learning to the denoising model about the known artifacts and the physical features of a material system. The FFT-based loss function helps to better minimize the random noise and artifacts which otherwise are not sufficient for pixel-based loss. The tuning of the weights of the information from two input channels helps to balance the learning, without overfitting towards artifacts/noise removal and feature preservation. Combining these three, we demonstrated GDM model over several microscopy (such as STM, AFM and SEM) images of different material systems, with different pattern of artifacts. We showcased the efficiency of the performance of GDM with other methods such as N2V and N2V2, without the FFT integration and with non-optimal tuning. We believe this proposed workflow can be extended to improve other microscopy image quality and will benefit the experimentalists with the proposed design flexibility to smartly tune via domain-experts' preferences. As in future scope, we aim to expand the architecture for automated and fast implementation to microscopy instruments for real-time denoising.

## Methods

**FFT loss calculation**

To quantitatively assess the structural similarity between the reconstructed and ground-truth images, we employed a Fourier-domain cosine similarity metric as FFT-loss function. The method transforms both the predicted and target image patches, $ip_{...}, \overline{ip_{...}}$, into the frequency domain using a two-dimensional Fast Fourier Transform (FFT). To ensure interpretability and consistency across spatial frequencies, the zero-frequency (DC) component is shifted to the center of the spectrum using *fftshift* function. Then, the absolute value of the shifted FFT spectra of the predicted and target image patches are computed. Finally, the quality score $\Psi_{FFT}$ is then computed as the cosine similarity between the FFT spectra vectors between the predicted and target patches as per **Eq. 4**:

$$\Psi_{FFT} = \frac{\langle |F_{\text{pred}}|, |F_{\text{target}}| \rangle}{(\||F_{\text{pred}}|\|_2 * \||F_{\text{target}}|\|_2) + \epsilon} \quad (4)$$

where $F_{\text{pred}}$ and $F_{\text{target}}$ denote the shifted FFT of the predicted and target images, respectively, and $\epsilon = 10^{-8}$ is a small numerical stabilizer. This metric provides a normalized score within $[0, 1 + \epsilon]$, where higher values indicate greater similarity in the frequency content between the reconstructed and true images.

**PNR score calculation for evaluation metric**

To quantitatively assess image quality of the original and denoised test images, we implemented an FFT-based peak analysis metric. The input image $I$ was first converted to grayscale and normalized to the range $[0,1]$. To suppress spectral leakage from sharp edges, a 2D Hann window was applied prior to transformation. The windowed image was transformed into the frequency domain using a two-dimensional Fast Fourier Transform (FFT) using *fft2* function in Numpy, and the zero-frequency (DC) component was centered via frequency shifting using *fftshift* function. Then, the absolute value and the log of the shifted FFT spectra of the image was computed, to enhance visibility of weak peaks. Lattice-related frequency peaks were detected using local maximum filtering function in *Skimage* package, peak_local_max, with parameter set as *min_distance* =10 and peak_prominence = 0.05, excluding the central DC peak to avoid bias from overall intensity. The mean power at the detected peak positions is computed as the **signal** $P_{\text{peak}}$, while the mean power in the high-frequency region (outside 50% of the maximum radius) is used as the **noise** $P_{\text{noise}}$.

The resulting Peak-to-Noise Ratio (PNR), expressed in decibels (dB) as per **Eq. 5**:

$$\text{PNR}_{\text{dB}} = 10\log_{10}\left(\frac{P_{\text{peak}}}{P_{\text{noise}}}\right) \tag{5}$$

A higher PNR value indicates a stronger, well-defined lattice periodicity and lower noise content in the images.

**Line-length score within specified angles for evaluation metric**

To quantitatively assess image quality of the original and denoised test images, we implemented the second evaluation metric to calculate total line-length within user specified angles (largely representing the artifacts). Here, to detect fine lines in the images, we implement a Canny edge detector function *canny*, followed by a probabilistic Hough transformation function *probabilistic_hough_line* in the *Skimage* package. The parameters of the *canny* function is set as $\sigma = 0.1$ (for STM images), $\sigma = 1$ (for AFM and SEM images), $threshold_{low} = 1$ (for STM and AFM images), $threshold_{low} = 5$ (for SEM images), $threshold_{high} = 10$. The parameters of the *probabilistic_hough_line* function is set as $threshold = 5$, $line_{length} = 1$, $line_{gap} = 2$ (for STM and AFM images), $line_{gap} = 1$ (for SEM images). Then, the length of each detected line is calculated, which then filters with the user specified angle range as $(\theta_1, \theta_2)$. Then we calculate the total length $\boldsymbol{L_{line}}$ as the sum of the length of lines within $(\theta_1, \theta_2)$ as per **Eq. 6**.

$$L_{line} = \sum_{\theta_l \subset (\theta_1, \theta_2)} \sqrt{(x_1 - x_2)^2 + (y_1 - y_2)^2} \tag{6}$$

**Preprocessing steps of STM training image of Cu (111)**

Here we provided the detailed steps to preprocess the Cu (111) STM training simulated and experimental images to enhance the weak but physically important features (ripples). The grayscale STM training images $I_{\text{sim}}, I_{\text{exp}}$ were first transformed into the frequency domain $F(I)$ using a two-dimensional Fast Fourier Transform using *fft2* function in Numpy, followed by frequency shifting using *fftshift* function. To selectively enhance surface ripples and suppress unwanted low- and high-frequency noise, a circular band-pass filter was constructed in the frequency domain as per **Eq. 7**:

$$M_r = \{(k_x, k_y): r_{low} < \sqrt{k_x^2 + k_y^2} < r_{high}\} \tag{7}$$

Where $M_r$ is the radial mask, $r_{low} = 20$ and $r_{high} = 60$ define the inner and outer radii in pixel units, $k_x$ and $k_y$ are the spatial frequency components along the x and y directions in the Fourier domain. Vertical streaks were further removed by excluding pixels within an angular margin ($\pm\theta = 30$) around the vertical frequency axis (90° and 270°), defined as angular mask $M_\theta$. The final mask was formed as the intersection of the radial and angular constraints:

$$M_{final} = M_r \cap M_\theta \tag{8}$$

Filtered frequency components were obtained as:

$$F_{filtered} = F \cdot M_{final} \tag{9}$$

and the enhanced real-space image was reconstructed using the inverse FFT functions such as *ifftshift* and *ifft2*, as per the shown in fig. 1:

$$I_{filtered} = F^{-1}(F_{filtered}) \tag{10}$$

The relevant python codes for this preprocessing function are generated and adapted from ChatGPT via domain-specific prompt engineering to enhance the concurrent circles of the STM images.

**Preprocessing steps of AFM training image of Pantoea sp. YR343 biofilm**

Here we provided detailed steps to preprocess the **Pantoea sp. YR343** biofilm AFM training experimental images to remove the horizontal scan lines. The grayscale AFM training image $I_{exp}$ was first transformed into the frequency domain $F(I)$ using a two-dimensional Fast Fourier Transform using *fft2* function in Numpy, followed by frequency shifting using *fftshift* function. Gaussian smoothing, using *gaussian_filter1d* function in *scipy* package with $\sigma = 0.1$ was applied to reduce spurious peaks. To prevent removal of low-frequency background information, pixels within a central radius ($r = 50$ pixels) around the DC component were excluded from further analysis. Prominent peaks in the column-wise energy profile, corresponding to vertical frequency components of horizontal streaks, were identified using a peak detection algorithm *find_peaks* in *scipy* package. A notch mask was constructed to suppress identified streak frequencies while retaining other image information. The mask, $M_{FFT}$, set the corresponding frequency columns to zero within a half-width of 1 pixel. Frequencies within the DC exclusion radius ($r = 50$ pixels) were preserved. The filtered frequency domain image was obtained and the inverse FFT reconstructed the intermediate cleaned image in real space, as per **Eq. 11.**

$$I_{filtered} = F^{-1}(F \cdot M_{FFT}) \tag{11}$$

Dark regions, defined by the *dark_threshold* as pixels below the 50[th] percentile of intensity in the cleaned image, were further processed to remove residual horizontal streaks. A binary "***dark-masked image***", $I_{masked}(x, y)$, as shown in the middle image of fig. 6a, was generated from the *dark_threshold*.

Next, to overlay the features in the "***dark-masked image***", we merged with the respective original image. Thus, pixels with intensity values greater than 128 in the "***dark-masked image***" correspond to scan-line artifacts are set with uniform white pixel values as 255. Pixels below this threshold are preserved to the intensity of the pixels of the original image. The merged image $I_{\text{merged}}(x, y)$ is generated as per **Eq. 12**, as shown in the bottom image of fig. 6a.

$$I_{\text{merged}}(x, y) = \begin{cases} 255, & I_{\text{masked}}(x, y) > 128 \\ I_{\text{orig}}(x, y), & \text{otherwise} \end{cases} \tag{12}$$

The relevant python codes for this preprocessing function are generated and adapted from ChatGPT via domain-specific prompt engineering to remove the horizontal scan lines of the AFM images.

**Preprocessing steps of SEM training image of Bacterial consumed Plastic degradation.**
Scanning electron microscopy (SEM) images often contain bright strip artifacts resulting from detector noise or scanning inconsistencies. Here we provided detailed steps to preprocess SEM training image to detect and remove these artifacts while preserving underlying image features. The original SEM image, denoted as $I_{\text{SEM}}$ was first converted to grayscale ($I_{\text{gray}}$). Then, bright strip artifacts were identified by thresholding the grayscale image using *threshold* function in *OpenCV* package. Pixels with intensity values above a chosen threshold ($T = 130$) were considered part of the artifact regions. This threshold was chosen after several iterations and validations from visualization in order to minimize the elimination of the non-artifact regions. A binary mask $M_{\text{strips}}$ was generated such that this mask captures regions with unusually high brightness corresponding to strip artifacts, as per **Eq. 13**.

$$M_{\text{strips}}(x, y) = \begin{cases} 1, & I_{\text{gray}}(x, y) > T \\ 0, & \text{otherwise} \end{cases} \tag{13}$$

The detected artifact regions were removed using an image inpainting algorithm *inpaint* in *OpenCV* package, finally to obtain the "***cleaned image***" as shown in bottom image of fig. 8a. Specifically, the *Telea* inpainting method was applied, which iteratively fills masked regions using the surrounding pixel information. Here, the inpainting radius is set as $r = 3$ pixels. The relevant python codes for this preprocessing function are generated and adapted from ChatGPT via domain-specific prompt engineering to remove the bright spots of the SEM images.

**Acknowledgements:**
This work (H.Z) was supported by the University of Tennessee startup funding of A.B. The authors (H.Z and A.B) acknowledge the use of facilities and instrumentation at the UT Knoxville Institute for Advanced Materials and Manufacturing (IAMM) and the Shull Wollan Center (SWC) supported in part by the National Science Foundation Materials Research Science and Engineering Center program through the UT Knoxville Center for Advanced Materials and Manufacturing (DMR-2309083). The STM experiment was supported by the National Science Foundation Materials Research Science and Engineering Center program through the UT Knoxville Center for Advanced Materials and Manufacturing (DMR-2309083) (C.V.), by the UT-Oak Ridge Innovation Institute (UT-ORII) through the UT-ORII SEED grant (L.B.), and by the University of Tennessee startup funding of W.K. (W.Y.). Work by R.M, S.R.B, M.C and L.C were supported by the U.S. Department of Energy, Office of Science FWP ERKCZ64, Structure Guided Design of Materials to Optimize the Abiotic-Biotic Material Interface, as part of the Bio preparedness Research Virtual Environment (BRaVE) initiative. AFM measurements and sample preparation were conducted as part of a user project at the Center for Nanophase Materials Sciences (CNMS), which is a US Department of Energy, Office of Science User Facility at Oak Ridge National Laboratory. The authors would like to sincerely thank Scott Retterer for his generous assistance in the acquisition of BRAVE funding for this research. Work by K.S.A and A.G were provided in part by the U.S. Department of Energy, Office of Energy Efficiency and Renewable Energy, Bioenergy Technologies Office (BETO) and Advanced Materials and Manufacturing Technologies Office (AMMTO) as part of the BOTTLE Consortium. Zeiss Merlin SEM was performed at the Center for Nanophase Materials Sciences, which is a U.S. Department of Energy Office of Science user facility at Oak Ridge National Laboratory (ORNL).

**Contributions**
A.B and W.K conceived and supervised the project. H.Z and A.B designed the GDM algorithm, wrote the codes for implementation and analysis, analyzed data, prepared figures. C.V, L.B and W.Y prepared STM datasets for model validation and assisted with analysis of Case Study 1. W.K supervised and supported funding for C.V, L.B and W.Y. R.M conducted AFM experiments to generate data, assisted with analysis of Case Study 2. M.C and L.C assisted with the AFM data preparation. S.R.B grew the bio-film samples used in Case Study 2. K.S.A prepared SEM datasets for model validation and assisted with analysis of Case Study 3. A.G supervised and supported funding for K.S.A. A.B wrote the manuscript while H.Z, C.V, L.B, W.K and K.S.A co-wrote the manuscript. All authors have reviewed and provided feedback on the manuscript.

**Conflict of Interest:**
The author confirms there is no conflict of interest.

**Code and Data Availability Statement:**
The analysis reported here along with the code is summarized in Notebook for the purpose of tutorial and application to other data and can be found in https://github.com/arpanbiswas52/paper-code-microscopyGDM